\documentclass[12pt]{article}
\usepackage{sectsty}
\usepackage[T1]{fontenc}
\usepackage{kpfonts}
\usepackage[dvipsnames]{xcolor}
\usepackage{pdfpages}
\usepackage{sgame}
\usepackage{accents}
\usepackage{tikz-cd}
\usepackage{float}
\usepackage[round]{natbib}
\usepackage{multirow}
\usepackage{dutchcal}
\usepackage{pdfpages}
\usepackage{bbm}
\usepackage{sgame}
\usepackage{mathtools}
\usepackage{amsthm}
\usepackage{enumitem}
\usepackage[margin=1in]{geometry}
\usepackage{caption}
\usepackage{subcaption}
\usepackage{epigraph}
\usepackage{listings}
\usetikzlibrary{calc}
\usetikzlibrary{shapes,arrows}
\usepackage{tcolorbox}

\newcommand{\ubar}[1]{\underaccent{\bar}{#1}}
\DeclareMathOperator\supp{supp}

\DeclareMathOperator*{\epi}{epi}

\DeclareMathOperator*{\argmax}{arg\,max}

\newtheorem{theorem}{Theorem}[section]
\newtheorem{proposition}[theorem]{Proposition}
\newtheorem{lemma}[theorem]{Lemma}
\newtheorem{corollary}[theorem]{Corollary}
\newtheorem{claim}[theorem]{Claim}

\theoremstyle{definition}

\newtheorem{definition}[theorem]{Definition}
\newtheorem{example}[theorem]{Example}

\newtheorem{remark}[theorem]{Remark}

\sectionfont{\color{BlueViolet}}
\subsectionfont{\color{BlueViolet}}

\usepackage{hyperref}
\hypersetup{
    colorlinks=true,
    linkcolor=CadetBlue,
    filecolor=CadetBlue,      
    urlcolor=CadetBlue,
    citecolor=Mulberry,
}

\usepackage{setspace}

\definecolor{backcolour}{rgb}{0.63, 0.79, 0.95}

\lstdefinestyle{mystyle}{
  backgroundcolor=\color{backcolour},
  basicstyle=\ttfamily\footnotesize,
  breakatwhitespace=false,         
  breaklines=true,                 
  captionpos=b,                    
  keepspaces=true,                 
  numbers=left,                    
  numbersep=5pt,                  
  showspaces=false,                
  showstringspaces=false,
  showtabs=false,                  
  tabsize=2
}

\providecommand{\keywords}[1]{\textbf{\textit{Keywords:}} #1}
\providecommand{\jel}[1]{\textbf{\textit{JEL Classifications:}} #1}

\begin{document}
\title{Bayesian Elicitation}
\author{Mark Whitmeyer\thanks{Arizona State University \newline Email: \href{mailto:mark.whitmeyer@gmail.com}{mark.whitmeyer@gmail.com}. Earlier versions of this paper were titled ``On Optimal Transparency in Signaling Games,'' and ``Opacity Design in Signaling Games.'' This paper also incorporates the previously separate working paper ``In Simple Communication Games, When Does \textit{Ex-ante} Fact-finding Benefit the Receiver?'' I thank V. Bhaskar, William Fuchs, Rosemary Hopcroft, Vasudha Jain, Meg Meyer, Vasiliki Skreta, Max Stinchcombe, Yiman Sun, Tymon Tatur, Alex Teytelboym, Thomas Wiseman, Joseph Whitmeyer, Kun Zhang, and various seminar and conference audiences for their feedback. An earlier version of this work was the third chapter of my dissertation at UT Austin. I worked on this while at the Hausdorff ($T_2$) Center for Mathematics in Bonn, supported under the DFG Project 390685813}}
\date{\today{}}% 
\maketitle
\begin{abstract}
We study how a decision-maker can acquire more information from an agent by reducing her own ability to observe what the agent transmits. In a large class of binary-action games, opacity design is just as good as full commitment to actions and also guarantees that \textit{ex ante} information acquisition always benefits the receiver, even though without opacity design this learning might actually lower the receiver's expected payoff. %These ideas are applied to a range of economically relevant scenarios.
\end{abstract}
\\
\jel{C72; D82; D83}

\newpage

\section{Introduction}

There is a decision-maker who must choose an action in an uncertain world. She does not have direct access to information about the state of the world, but there is a second agent who does. The second agent (or sender) observes the state before sending a message\footnote{We term the sender's action a message in order to distinguish it from the receiver's action. In some settings, like for instance cheap talk games, the moniker is fitting. In others, like for instance the \cite{spence} model of signaling through education attainment, labeling his action a message is less appropriate.}, which the decision-maker (or receiver) observes before taking an action. This scenario is called a signaling game (or a communication game), and the class of such games includes cheap-talk games, like that studied in \cite{cs}; and costly signaling games, like that studied in \cite{spence}.

In many situations, the receiver's only concern is information transmission: she values only the information content of the message, and so her welfare increases as the sender becomes more informative. However, there are a number of potential frictions that could impede this communication. First, the sender's and the receiver's preferences over the action taken may be imperfectly aligned. Second, the messages may be costly to the sender, with these costs affecting the messages that are chosen in equilibrium. As a result, less than full information may be transmitted at equilibrium, and the frictions may even be so severe that no information is transmitted.

Now, suppose that the receiver is not forced to observe the sender's message directly, but may commit \textit{ex ante} to observe a noisy signal of the message instead. Can such opacity design help the receiver? How should the receiver design an optimal information structure?

If this were a decision problem in which the information were exogenous, then the answer to the first question would be no. The receiver would always (at least weakly) prefer to observe the message itself rather than some noisy signal. Here, however, the message is not exogenous but is instead an equilibrium choice of the sender. The sender is aware of the information structure, which therefore affects the message that he sends at equilibrium. Consequently, a noisier signal of the sender's message may beget a more informative message, to the receiver's gain.

%There is an important trade-off inherent to opacity design. By choosing a more informative signal of the message, the receiver obtains more information for any fixed strategy of the sender. On the other hand, the sender's strategy--and hence the information--is an endogenous choice that he makes cognizant of the information structure. Thus, a noisier signal of the sender's message may beget a more informative message, to the receiver's gain.

The classic ``Beer-Quiche" example of \cite{cho}, reproduced in Figure \ref{fig}, provides a useful illustration of this idea. The sender is privately informed about whether he is wimpy or strong ($\theta_{W}$ or $\theta_{S}$, respectively). His message is his choice of fare, either beer ($B$) or quiche ($Q$). All else equal, wimps prefer quiche and strong guys beer, but there are strategic concerns as well: the receiver observes the sender's order before choosing whether to fight him. The receiver prefers to fight the wimp and not fight the strong guy, but the sender does not wish to be fought, regardless of the state. Moreover, in each state, the sender would rather consume his least favorite fare and not be fought than consume his favorite fare and be fought. Both sender and receiver share the common prior $\mu_{0} \coloneqq \Pr\left(\theta_{S}\right) = 2/3$: the sender is less likely to be the wimp.

\begin{figure}
    \centering
    \begin{tikzpicture}[scale=1.4,font=\footnotesize]
\tikzset{
solid node/.style={circle,draw,inner sep=1.5,fill=black},
hollow node/.style={circle,draw,inner sep=1.5}}
\tikzstyle{level 1}=[level distance=12mm,sibling distance=25mm]
\tikzstyle{level 2}=[level distance=15mm,sibling distance=15mm]
\tikzstyle{level 3}=[level distance=17mm,sibling distance=10mm]
\node(0)[hollow node]{}
child[grow=up]{node[solid node,label=above:{$\theta_{W}$}] {}
child[grow=left]{node(1)[solid node]{}
child{node[solid node,label=left:{$(0, 1)$}]{} edge from parent node [above]{$F$}}
child{node[solid node,label=left:{$(2, 0)$}]{} edge from parent node [below]{$NF$}}
edge from parent node [above]{$B$}}
child[grow=right]{node(3)[solid node]{}
child{node[solid node,label=right:{$(3, 0)$}]{} edge from parent node [below]{$NF$}}
child{node[solid node,label=right:{$(1, 1)$}]{} edge from parent node [above]{$F$}}
edge from parent node [above]{$Q$}}
edge from parent node [right]{$1-\mu_0$}}
child[grow=down]{node[solid node,label=below:{$\theta_{S}$}] {}
child[grow=left]{node(2)[solid node]{}
child{node[solid node,label=left:{$(1, 0)$}]{} edge from parent node [above]{$F$}}
child{node[solid node,label=left:{$(3, 1)$}]{} edge from parent node [below]{$NF$}}
edge from parent node [above]{$B$}}
child[grow=right]{node(4)[solid node]{}
child{node[solid node,label=right:{$(2, 1)$}]{} edge from parent node [below]{$NF$}}
child{node[solid node,label=right:{$(0, 0)$}]{} edge from parent node [above]{$F$}}
edge from parent node [above]{$Q$}}
edge from parent node [right]{$\mu_0$}};
\draw[dashed,rounded corners=10]($(1) + (-.45,.45)$)rectangle($(2) +(.45,-.45)$);
\draw[dashed,rounded corners=10]($(3) + (-.45,.45)$)rectangle($(4) +(.45,-.45)$);
\node at ($(1)!.5!(2)$) {$R$};
\node at ($(3)!.5!(4)$) {$R$};
\end{tikzpicture}
    \caption{The Parameterized ``Beer-Quiche" Game}
    \label{fig}
\end{figure}

If the receiver perfectly observes the sender's message, there are no equilibria in which any information is transmitted. The logic behind this is simple: there can be no equilibrium in which both $B$ and $Q$ are sent in such a way that the receiver strictly prefers to take a different action after each message. In such a circumstance, the wimp always prefers to deviate to the message that is followed by not fight. The receiver's payoff is $2/3$.

Suppose the scenario is amended to include a neutral third party, a waiter. The receiver no longer observes the sender's message directly; instead, the waiter witnesses the message before communicating to her. Because the waiter is neutral, he can be represented as a signal, conditional distributions $\pi\left(\cdot|B\right)$ and $\pi\left(\cdot|Q\right)$ on some set of signal realizations.\footnote{Throughout this paper, the signal, $\pi$, is conditioned on the message, $m$, and not the state, $\theta$.} Such obfuscation can strictly increase the receiver's payoff. The optimal signal is
\[\begin{split}
\pi\left(f|Q\right) &=  \frac{1}{2}, \qquad \pi\left(f|B\right) = 0\\
\pi\left(nf|Q\right) &=  \frac{1}{2}, \qquad \pi\left(nf|B\right) = 1
\end{split} \text{ .}\]

After $nf$ the receiver does not fight and after $f$ the receiver fights. The signal begets a separating equilibrium--one in which the wimp chooses $Q$ and the strong guy chooses $B$. This gain in informativeness outweighs the garbling by the signal, and the receiver obtains a strictly higher equilibrium payoff than without opacity ($5/6$ versus $2/3$).

%In these three examples, the role of the signal is played by a neutral third party (mediator). In other cases, the signal may correspond to coarse accounting or disclosure rules. Consider for instance a venture-capital environment, in which the equity-retainment decisions by entrepreneurs signal their private information (\cite{lp}). A prospective investor may benefit by only allowing the investor to disclose only partial (or certain kinds of) information about her stake.  %Finally, as explored in \cite{whit2}, partial inattention toward her offspring's cries can benefit a parent who wishes to donate resources only if her child is in need.

Throughout this paper, we explore the problem of solving for the receiver-optimal\footnote{We focus throughout on the receiver-optimal Perfect Bayesian Equilibrium.} information structure in signaling games. We restrict attention to the environment in which the sender and receiver share a common prior and focus on signaling games in which the receiver has no intrinsic preferences over the message chosen by the sender.%The ``Beer-Quiche'' example is one such game: the receiver is indifferent about the sender's order outside of the information that the choice contains. 

%Opacity design is difficult. The receiver solves an optimization problem with two sets of constraints: given the signal, the receiver's action must be sequentially rational and the sender's strategy must be optimal. This is a maximization problem with incentive compatibility constraints for the sender in each state and obedience constraints for the receiver. 

There is a useful upper bound for the solution to the opacity design problem; the \textit{commitment} solution for the receiver, which corresponds to the scenario in which the receiver can commit to a distribution over actions conditioned on the sender's choice of message. Consequently, if the commitment strategy that maximizes the receiver's payoff also satisfies the receiver's obedience constraints, then it must correspond to the optimal information structure. The commitment problem is much simpler, and we establish that it reduces to a linear program.

In the first main result of this paper, Proposition \ref{bigiftrue2}, we discover that in any signaling game with two actions, ``opacity equals commitment”. Namely, for any number of states and messages, provided the receiver has a binary decision, the commitment solution satisfies the receiver's obedience constraints.

We also investigate the benefits and drawbacks to the receiver of \textit{ex ante} information acquisition, and how the consequences thereof are affected by the receiver's ability to design the opacity in the ensuing game. While it is clear that in any setting, information acquisition may benefit the receiver, our goal is to discover in which settings \textit{ex ante} learning is always beneficial. Surprisingly, even in games with just two actions, \textit{ex ante} information acquisition may hurt the receiver if she may not design the opacity in the ensuing game. However, the second main result of this paper, Proposition \ref{convconv}, establishes that opacity design flips this result. In two-action games, the value of information is always (weakly) positive, provided the receiver may design the opacity in the game that follows.

\subsection{Discussion of Assumptions and Relevance}

Let us briefly discuss the relevance of this paper's theoretical environment.

\subsubsection{Real-World Instances of Opacity Design} There are coarse ratings or grades, in which case the signal is the policy that maps the sender's behavior to ratings or grades; recommender or aggregation systems, in which case the signal is the aggregator or recommender who distills product information like warranty details, return policies, advertising and even price details,\footnote{These are all classic signaling environments; see \cite{mil, nelson} and \cite{gal}.} into a purchase recommendation for a consumer; and interactions through third parties (mediators), in which case it is the mediator itself that serves as the signal. The ``Beer-Quiche'' game discussed above can be viewed as an allegory for an incumbent firm signaling to a potential entrant through its pricing decisions. Accordingly, we argue a role for a third party consultant who advises the aspiring entrant. Transparency is often lauded as a desirable feature in politics. However, political decision makers have private information, so their actions can carry information. Our framework shows how a neutral free press that in some circumstances obfuscates and hides an incumbent's actions may benefit a populace.

\subsubsection{Who Are These Mediators?} One profession that arguably fills this role is headhunting or staffing firms. There are many non-productive activities that prospective job candidates take part in that are nevertheless useful to a hiring firm due to the information these behaviors carry about the hidden productivity of the candidates. In \cite{spence} it is education that is such a (intrinsically) wasteful activity. This paper, therefore, illustrates the usefulness of hiring through third party agencies, who serve as intermediaries and thereby improve welfare even if they do not possess any special expertise in identifying talent.

\subsubsection{Focusing on the Receiver's Design Problem.}

In a number of examples, it seems realistic to have the receiver design the mechanism. For instance, consider a university that is faced with the problem of admitting students. It is the school that designs the admissions form, which is effectively the opacity-design problem under study in this work: the university can allow the student to enter his standardized test school in full detail, only a coarse categorization of the score, or no information about the score. Naturally, the university does this to optimize its (the receiver's) welfare. Similarly, in the staffing-firm and consulting examples mentioned above, it is the receiver--hiring firm and potential entrant, respectively--that typically pays the staffing firm (or consultant) for its service. Given this, the intermediary (in order to maximize its own rents) designs the mechanism to be receiver-optimal.\footnote{Moreover, it is in the intermediary's interest to be \textit{unbiased} at an interim point, as otherwise its services are worthless. This can be sustained, for instance, through reputation.}

\subsubsection{Relevance of Two Actions.}

Our main results are based on a binary-action specification, which may be limiting in some environments, as receivers may have more than two available actions. However, many important scenarios \textit{do} only have two options. For instance, a university admits or rejects an applicant, a populace keeps or ousts an incumbent, and a consumer purchases or refrains from buying a product. Additionally, a hiring firm either hires or rejects an applicant, and a firm considering market entry chooses whether to enter or stay out of the market.

\subsection{Related Work}

The idea that introducing noise to communication can enhance welfare is not new to this paper. In particular, a number of papers study this in various cheap talk settings. \cite{for1} explores mediation in a (cheap-talk) job-market example; and \cite{my91} describes a game in which communication is improved if occasionally messages are not transmitted.\footnote{Amusingly, the medium for communication is a (somewhat wayward) carrier pigeon.} \cite{blume} shows how noise can improve welfare in a uniform-quadratic setting, and other papers follow--\cite{gol,ivan}; and \cite{ray}--that look at the benefits of mediation in the uniform-quadratic environment. More recently, \cite{sal} explores a general version of optimal mediation using duality theory from linear programming.

\cite{gol} and \cite{blume} focus on receiver-optimal equilibria, as do we in this paper.\footnote{\cite{ivan} and \cite{ray} investigate slightly different questions within the uniform quadratic framework, rather than tackle a design problem.} \cite{sal}, in contrast, looks at the design problem from the sender's perspective. In that light, his paper is more true to the standard Bayesian persuasion paradigm, in which a sender produces information to influence a receiver. Here, we tackle an extraction (or elicitation) problem instead: how can a receiver induce a sender to reveal information to her?

Because the messages are cheap talk, these papers view the mediation problem as a centralized problem and use the \textit{Communication Equilibrium} concept, as formulated in \cite{my86} and \cite{for2}. Here, provided the game is cheap talk and the set of messages is sufficiently large, the design of the optimal signal, $\pi$, is equivalent to the centralized mediator-driven problem analyzed in these cheap talk papers (we can think of the separating equilibrium as the reporting of the state). However, if the message set is not big enough, then this equivalence is lost. Moreover, if the messages are costly (i.e., the game is not cheap talk), then the problem we explore is different from the centralized problems of the literature. With costly messages, it is not even without loss of generality to restrict the sender to pure strategies.% (deterministic strategies from states to messages).

It is more difficult to obtain general results when messages are not cheap. In his exploration of signaling games, \cite{rick}'s main result is that welfare is improved by a signal only if either i. the signal removes some profitable deviations to unused messages and/or ii. the signal garbles the distribution over posteriors induced by the sender's strategy non-trivially. \cite{ball} investigates the design of a scoring rule in order to elicit information from a sender who can distort multiple features (what the receiver can observe) about himself. Analogously, he finds that coarser information can benefit the receiver due to the endogeneity of the information, which is produced by the sender. In a similar setting, \cite{frank} look at the commitment problem for a principal acquiring manipulable information from an agent. %Ball's paper echoes the trade-off that we consider here, that, ``the intermediary must consider how the scoring rule motivates the sender to distort her features.'' Thus, a noisier signal of the sender's message may beget a more informative message, to the receiver's gain.

In contrast to these papers, this paper highlights the power of opacity design in signaling games. First, it is just as good for the receiver as full commitment; and second, it guarantees that the value of \textit{ex ante} information is always positive. This first equivalence is especially useful: it transforms a large and difficult problem into a much easier linear program. It is not uncommon for applied papers to include a commitment benchmark, and here we find that this benchmark can be obtained through information alone.

The equivalence of opacity design and commitment also sheds light on how increased noise can improve the receiver's welfare. Because the obedience constraints are always satisfied by the commitment solution, the receiver's interim incentives (at the optimum) are inconsequential. Accordingly, maximizing the receiver's payoff is constrained only by the sender's incentives.

%The second point noted above--that opacity design ensures that \textit{ex ante} learning cannot hurt the receiver--is also important, and contrasts with the other results of Section \ref{convexity}, where we find that in opacity design's absence, information acquisition may be to the receiver's detriment, even when she has just two actions. %That section also contains novel results on the value of information in signaling games, and presents (tight) sufficient conditions that guarantee \textit{ex ante} learning benefits the receiver.

This paper is also related to the recent line of information design papers in which the information generation process (or even the state) is endogenous. Those papers include \cite{bol}, who study a persuasion problem in which the endogenous state depends on the hidden effort of a third party; \cite{szentes}, who ask how to optimally monitor an agent; and \cite{afg}, who investigate how a planner should reveal information about past trading volume. The closest of these to this work is \cite{bol}, whose problem reduces formally to a constrained persuasion problem, the constraints to which correspond to the incentive compatibility conditions of the effort-providing agent. The opacity design problem is a constrained persuasion problem, and we use the Lagrangian approach introduced by \cite{bol} (see also \cite{dov}) to solve the game of \(\S\) \ref{mtbgame}. %To help deal with some technical issues related to this idea of a ``constrained information design'' problem, several authors have written notes, see Doval and Skreta (2018) \cite{dov} and Zhong (2018) \cite{zhong}.

Finally, this paper is also related to the rational inattention literature, which explores the effects of costly information processing and/or acquisition on an agent's (or agents') behavior. \cite{macksurvey} provides an excellent survey of this area. 
At a fundamental level, our receiver can be viewed as solving a rational inattention problem, where the menu of feasible information acquisition strategies is determined by the various incentive constraints (obedience and sender incentive compatibility).\footnote{I thank a referee for pointing out this perspective.} In contrast to the literature; however, in which more information is typically assumed to be more expensive, this effective cost of ``acquiring'' information can take unusual forms. For instance, a game that admits a separating equilibrium but not equilibria with certain varieties of partial state revelation would be such that full information (learning the state) is costless for the receiver, but some garblings of the prior are infinitely costly.

%A recent paper by Denti suggests a potentially profitable future direction that incorporates ideas from this paper as well as the rational inattention literature. He studies signaling games in which the receiver is rationally inattentive and chooses a signal of the sender’s message subject to a cost (again, at the interim stage). Using this he develops a novel refinement: what equilibria are selected as the receiver’s attention cost vanishes?

\section{The Model}\label{bq}

The setup is a version of the standard signaling game. There are two players: a sender, $S$; and a receiver, $R$. The sender is privately informed about the state  $\theta \in \Theta$  and chooses a message, $m$, from the set of messages $M$. The receiver observes $m$, but not $\theta$, updates her belief about the state using Bayes' law, then chooses a mixture over actions $a$ from the set of actions $A$. We assume that sets $A$ and $M$ are compact and that $\Theta$ is finite.

$S$ and $R$ share a common prior over the state of the world, $\mu \in \Delta(\Theta)$, where $\mu(\theta) \coloneqq \Pr(\theta)$. Each player, $S$ and $R$, has state-dependent preferences over the message sent, and the action taken, which are represented by the continuous utility functions $u_{S}\colon A \times M \times \Theta \to \mathbb{R}$ and $u_{R}\colon A \times \Theta \to \mathbb{R}$. A behavioral strategy for $S$, $\sigma(\cdot | \theta)$, is a family of probability distributions over $M$. Similarly, a behavioral strategy for $R$, $\rho(\cdot | m)$, is a family of probability distributions over $A$. The special case in which messages are costless and the message space is large is the classical cheap-talk setting. We focus on receiver-optimal Perfect Bayesian Equilibria (PBE), defined in the standard manner.\footnote{Strictly speaking, we focus on receiver-optimal Perfect Bayesian Equilibria of the extended game in which the receiver commits \textit{ex ante} to a garbling of the sender's message, $\pi$.}

\subsection{Opacity Design}

Suppose that $R$ can design the game's opacity by committing \textit{ex ante} to observe the realization of a signal $\pi$, rather than $m$. A signal, $\pi$, is a mapping $\pi\colon M \to \Delta\left(X\right)$, where $\pi\left(x|m\right) \coloneqq \Pr\left(X = x | M=m\right)$ and $X$ is a (large) finite set of signal realizations. The timing is as follows: first, $R$ commits (publicly) to $\pi$; second, $S$ mixes over actions according to $\sigma\left(\cdot \vert \theta\right)$; and third, $R$ observes the signal realization $x$, before choosing a behavioral strategy $\rho\colon X \to \Delta(A)$.

The receiver solves
\[\sup_{\sigma, \pi, \rho}\mathbb{E}_{\mu, \sigma, \pi, \rho}\left[u_{R}\left(a, \theta\right)\right] \text{ ,}\]
such that
\[\label{Obedience}\tag{$O_{x}$}
\rho(a|x) \in \argmax_{\rho(a|x)}\mathbb{E}_{\mu, \sigma, \pi, \rho}\left[u_{R}\left(a,\theta\right)|x\right] \text{ ,}\]
for all $x \in X$;\footnote{Note that following any signal realization, the receiver will have beliefs concerning the sender's message and the state. Concordant with our focus on PBE, these beliefs must be consistent with Bayes' law whenever possible.} and
\[\label{ICi}\tag{$IC_{\theta}$}\sigma\left(m|\theta\right) \in \argmax_{\sigma\left(m|\theta\right)}\mathbb{E}_{\sigma, \pi, \rho}\left[u_{S}\left(a,m,\theta\right)|\theta\right] \text{ ,}\]
for all $\theta \in \Theta$.

The second group of conditions (Conditions \ref{ICi}) collects the optimality conditions for the sender typically found in signaling games (with noise): given the (expected) best responses by the receiver to a sender's message, in each state the sender's message must be optimal. The first set of conditions (Conditions \ref{Obedience}) collects the obedience constraints for the receiver, which mandate that her choices following each signal realization $x$ are sequentially rational. The term ``obedience'' hints at the first result, in which we note that a revelation principle applies. This allows us to greatly simplify the receiver's problem by restricting attention to direct signals that recommend actions. %First, a definition:

\begin{definition}
Call a signal $\pi$ \hypertarget{direct}{\textbf{Direct}} if $X = A$. That is, a direct signal recommends actions.
\end{definition}

Then, the following remark establishes that it is without loss of generality to restrict attention to direct signals, which recommend actions to the receiver. 

\begin{remark}\label{revp}
For any equilibrium triple, $(\sigma, \pi, \rho)$, there is another equilibrium triple, $(\sigma', \pi', \rho')$, that yields the same (terminal) joint distribution over messages and actions, and hence the same payoffs; where $\pi'$ is a direct signal, $\sigma = \sigma'$, and $\rho'(a|a') = 1$ for $a' = a$ and $\rho'(a|a') = 0$ for $a' \neq a$.
\end{remark}

In the remainder of the paper, we restrict attention to direct signals, $\pi$, and thus the receiver's problem can be reduced to
\[\sup_{\sigma, \pi}\mathbb{E}_{\mu, \sigma, \pi}\left[u_{R}\left(a,\theta\right)\right] \text{ ,}\]
such that
    \[\label{Obediencer}\tag{$O_{a}$}\mathbb{E}_{\mu, \sigma, \pi}\left[u_{R}\left(a,\theta\right)|a\right] \geq \mathbb{E}_{\mu, \pi, \sigma}\left[u_{R}\left(a',\theta\right)|a\right] \text{ ,}\]
for all $a$, $a' \in A$; and \[\label{ICir}\tag{$IC_{\theta}$}\mathbb{E}_{\pi}\left[u_{S}\left(a,m,\theta\right)|\theta\right] \geq \mathbb{E}_{\pi}\left[u_{S}\left(a,m',\theta\right)|\theta\right] \text{ ,}\]
for all $m \in \supp \sigma\left(\cdot|\theta\right)$, for all $m' \in M$, for all $\theta \in \Theta$. We term this program the \textbf{Opacity Design Problem}, and label a pair $\left(\sigma, \pi\right)$ that solves this program the \textbf{Opacity Design Solution}. Define $V = V\left(\mu\right)$ to be the value of this constrained optimization problem as a function of the prior. We call this the receiver's \textbf{Opacity Design Payoff}.

It is important to note that if one were to restrict the \textit{sender} to a pure strategy in each state, one would sacrifice generality. That is, the opacity design solution may require the sender to mix in some states (which is easy to show through a counterexample). However, as noted earlier, if the game is cheap talk (and there are at least as many messages as states) then the sender may be assumed to ``report'' the state in her message and hence play a pure strategy in the opacity design solution (this is just the Revelation Principle).   %Refer to Proposition 2.4 in an earlier draft of this work, \cite{Whit4}, which proves this result by means of

\section{Opacity Design = Commitment}\label{simple}

If the receiver has the stronger form of commitment to actions--i.e., instead of choosing a signal $\pi$, she chooses a conditional distribution over actions as a function of the sender's message, (with abuse of notation) $\pi$.\footnote{We are merely replacing ``recommendations'' with ``orders.''} This program, which we term the \textbf{Commitment Problem} is 
\[\sup_{\sigma, \pi}\mathbb{E}_{\mu, \sigma, \pi}\left[u_{R}\left(a,\theta\right)\right] \text{ ,}\]
such that
\[\label{ICcomm}\tag{$IC_{\theta}$}\mathbb{E}_{\pi}\left[u_{S}\left(a,m,\theta\right)|\theta\right] \geq \mathbb{E}_{\pi}\left[u_{S}\left(a,m',\theta\right)|\theta\right] \text{ ,}\]
for all $m \in \supp \sigma\left(\cdot|\theta\right)$, for all $m' \in M$, for all $\theta \in \Theta$. We label a pair $\left(\sigma, \pi\right)$ that solves this problem the \textbf{Commitment Solution}. The receiver's value is her \textbf{Commitment Payoff}.

Since the receiver's commitment problem is merely the opacity design problem absent the obedience constraints, the following observation is evident: 
\begin{remark}
Any (terminal) joint distribution over messages and actions that can be obtained through opacity design can be obtained through commitment to actions.
\end{remark}

As the next result illustrates, the commitment problem is much easier to solve than the opacity design problem, since we may restrict the sender to pure strategies in each state. Indeed, the result is trivial: the receiver's objective is linear in the probability of any message in the support of a sender's mixed strategy; and in any state in which the sender is mixing, he must be indifferent over all pure strategies in the support of his mixed strategy. Formally,

\begin{lemma}\label{lemlem}
There is a commitment solution in which the sender chooses a pure strategy in each state.
\end{lemma}

One consequence of this result is that there is a commitment solution for the receiver in which at most $|\Theta|$ messages are used. Furthermore, for a fixed vector of pure strategies, the receiver's commitment problem is a linear program. Thus, the receiver's commitment problem is merely a finite collection of linear programming problems. This also immediately implies existence of an optimum in the commitment problem.

Next, we discover the first main result:

\begin{proposition}[Opacity Design Equals Commitment]\label{bigiftrue2}
If $|A| = 2$, the opacity design solution coincides with the commitment solution.%the commitment solution for the receiver requires at most $|\Theta| = n$ signals and
\end{proposition}

\begin{proof}

Recall that in the commitment problem it is without loss of generality to restrict the sender to pure strategies in each state. Thus, the receiver's commitment payoff simplifies to
\[\int_{\theta \in \Theta}\left(\pi_{\theta}v_{\theta} + \left(1-\pi_{\theta}\right)w_{\theta}\right)d\mu(\theta) \text{ ,}\]
where $v_{\theta} \coloneqq u_{R}\left(a_{1}, \theta\right)$, $w_{\theta} \coloneqq u_{R}\left(a_{2}, \theta\right)$ and $\pi_{\theta} \coloneqq \pi\left(a_{1}|m_{\theta}\right)$. Without loss of generality, the receiver's optimal action under the prior is $a_{1}$. Accordingly,
\[\int_{\theta \in \Theta}\left(\pi_{\theta}v_{\theta} + \left(1-\pi_{\theta}\right)w_{\theta}\right)d\mu(\theta) \geq \int_{\theta \in \Theta}\left(\pi_{\theta}v_{\theta} + \left(1-\pi_{\theta}\right)v_{\theta}\right)d\mu(\theta) \geq \int_{\theta \in \Theta}\left(\pi_{\theta}w_{\theta} + \left(1-\pi_{\theta}\right)w_{\theta}\right)d\mu(\theta)  \text{ ,}\]
where the first inequality uses the fact that the receiver's commitment payoff must be weakly higher than her payoff from taking the action that is optimal under the prior and the second inequality follows by the optimality of $a_{1}$ under the prior.

This chain of inequalities implies that the two obedience constraints are satisfied:
\[\label{oo1}\tag{$O1$} \int_{\theta \in \Theta}\pi_{\theta}v_{\theta}d\mu(\theta) \geq \int_{\theta \in \Theta}\pi_{\theta}w_{\theta}d\mu(\theta)\text{ ,}\]
\[\label{oo2}\tag{$O2$}\int_{\theta \in \Theta}\left(1-\pi_{\theta}\right)w_{\theta}d\mu(\theta) \geq \int_{\theta \in \Theta}\left(1-\pi_{\theta}\right)v_{\theta}d\mu(\theta) \text{ .}\]
\end{proof}
%This result implies that starting with any signaling game, it is always possible to remove sufficiently many actions to yield opacity equals commitment. One one hand, it is possible that no actions need to be removed, even if the receiver's action set is large. This is true, for instance, if there exists a fully separating equilibrium in the game without opacity design. As the next section illustrates; however, two may be the maximal number of actions for which opacity equals commitment.
%\begin{corollary}\label{maximalcorr}
 %In any signaling game, there exists a number \(T \geq 2\) such that if \(\left|A\right| \leq T\) opacity \(=\) commitment. \end{corollary}

\subsection{Multiple Senders}

In some sense, the insight provided by Proposition \ref{bigiftrue2} is quite general: in fact, although this paper focuses two-player signaling games, opacity equals commitment for costly signaling games with an arbitrary number of senders provided the receiver has only two actions. To put differently, from the receiver's (principal's) perspective, \textit{information design is just as good as mechanism design (without transfers).} The exact same logic from the proof of Proposition \ref{bigiftrue2} goes through (though it is more cumbersome notationally since senders may be mixing): the receiver cannot strictly benefit from deviating at the interim stage (following a recommendation) since that would contradict \textit{ex ante} optimality. 

A special case of such a multi-sender environment is the classic mechanism design without transfers setting. There, the binary-action case is of particular relevance since it pertains to a principal's (in our parlance, receiver's) decision to approve a policy or candidate or convict a defendant--see for instance \cite{wolinsky2002eliciting}, \cite{feddersen2015persuasion}, \cite{battaglini2017public}, \cite{gradwohl2018persuasion}, \cite{ali2019should}, or \cite{kattwinkelwinter}, all of whom center their analysis around a principal's ability to commit to a binary decision. Our results imply that the strong form of commitment (to actions) assumed in these papers can be weakened to one of commitment to information, without affecting the receiver-optimal outcome.

Formally, suppose that rather than there only being one sender, there are \(k \geq 1\) (\(k \in \mathbb{N}\)) senders \(S_i\) (\(i = 1, \dots, k\)), each of which simultaneously sends a message from her (possibly idiosyncratic) set of messages \(M_i\). Each sender is informed of her dimension of the multidimensional state \(\theta_i \in \Theta_i\) and chooses a behavioral strategy \(\sigma_i\left(\left.\cdot\right| \theta_i\right)\), which is a probability distribution over messages. For this subsection alone, we redefine \(M \coloneqq \times_i M_i\), \(\Theta \coloneqq \times_i \Theta_i\), and \(\sigma \coloneq \left(\sigma_1, \dots, \sigma_k\right)\). Sender \(i\)'s utility function is \(u_{S_i} \colon A \times M \times \Theta \to \mathbb{R}\) and the receiver's utility is \(u_{R}\colon A \times \Theta \to \mathbb{R}\). All are continuous. \(\mu\) denotes the common prior and \(\mu_{-i}\) the joint distribution over states other than \(\theta_i\), given state \(\theta_i\). As is standard, let \(\sigma_{-i}\) (\(m_{-i}\)) denote the vector of mixed (pure) strategies chosen by senders other than \(i\).

As in the single-sender setting, the receiver commits \textit{ex ante} to a signal, mapping \(\pi \colon M \to \Delta \left(A\right)\), where we have used an analog of Remark \ref{revp} to allow us to restrict attention to direct signals. The receiver's opacity design problem is now
\[\sup_{\sigma, \pi}\mathbb{E}_{\mu, \sigma, \pi}\left[u_{R}\left(a,\theta\right)\right] \text{ ,}\]
such that
    \[\label{Obediencem}\tag{$O_{a}$}\mathbb{E}_{\mu, \sigma, \pi}\left[u_{R}\left(a,\theta\right)|a\right] \geq \mathbb{E}_{\mu, \pi, \sigma}\left[u_{R}\left(a',\theta\right)|a\right] \text{ ,}\]
for all $a$, $a' \in A$; and \[\label{ICm}\tag{$IC_{\theta_{i}}$}\mathbb{E}_{\pi, \mu_{-i}, \sigma_{-i}}\left[u_{S_i}\left(a,m_i,m_{-i},\theta_{i}\right)|\theta_{i}\right] \geq \mathbb{E}_{\pi, \mu_{-i}, \sigma_{-i}}\left[u_{S_i}\left(a,m_i',m_{-i},\theta_{i}\right)|\theta_{i}\right] \text{ ,}\]
for all $m_i \in \supp \sigma_i\left(\cdot|\theta_i\right)$, for all $m_i' \in M_i$, for all $\theta_i \in \Theta_i$, for all \(i = 1, \dots k\). As in the single-sender case, the commitment problem is just this program without the collection of obedience constraints (\ref{Obediencem}).

An exact analog of Proposition \ref{bigiftrue2} holds:
\begin{proposition}[Opacity Design Equals Commitment]\label{bigiftrue3}
With \(k \geq 1\) senders, if $|A| = 2$, the opacity design solution coincides with the commitment solution.%the commitment solution for the receiver requires at most $|\Theta| = n$ signals and
\end{proposition}
We omit the proof as it is identical to the proof of Proposition \ref{bigiftrue3} modulo some minor notational changes.

\section{A ``Match-the-Binary-State'' Game}

Let us briefly explore a simple binary-state game in order to i. investigate how the sender's and receiver's conflict of interest affects the optimal signal, ii. study the limits of ``Opacity \(=\) Commitment,'' and iii. illustrate how to solve the opacity design via Lagrangian methods.

Throughout this section, we impose that there are just two states, $\Theta = \left\{0,1\right\}$; and the set of actions for the receiver is a subset of $\left[0,1\right]$. The receiver's payoff, given action \(a\) and realized state \(\theta\), is the well-known quadratic loss function: \[U_R\left(a,\theta\right) = -\left(a -\theta \right)^2\text{ .}\] The sender has just two messages $M = \left\{0,1\right\}$. In addition, he has bias $b \in \left[\frac{1}{2},1\right)$. We also stipulate that the sender suffers a ``lying cost'' of $c \in \left[0, 2b-1\right)$. His payoff is, therefore, \[
    U_S\left(a,\theta, m\right) = \begin{cases} -\left(a -\left(\theta+b\right) \right)^2 \quad &\text{if} \quad m = \theta\\
    -\left(a -\left(\theta+b\right) \right)^2 - c \quad &\text{if} \quad m \neq \theta
\end{cases} \text{ .}\]

We begin by assuming the receiver has just two actions, in which environment Proposition \ref{bigiftrue2} holds. We then add one additional action and show that it fails to hold in general before, finally, solving a continuum of actions version of the game.

\subsection{Two Versus Three Actions}

First, suppose $A = \left\{0,1\right\}$. By Proposition \ref{bigiftrue2}, we need only solve the commitment problem for the receiver: she commits to two conditional distributions over actions $p \coloneqq \mathbb{P}\left(\left.1\right|1\right)$ and $q \coloneqq \mathbb{P}\left(\left.1\right|0\right)$. By the revelation principle (for scenarios with reporting costs, see \cite{conitzer}) it is optimal for the receiver to have the sender choose message $1$ when the state is $1$ and message $0$ otherwise. Thus, the receiver solves the program (where we have omitted constants)
\[\max_{p, q}\left\{-\mu \left(1-p\right) - \left(1-\mu\right)q\right\}\text{ ,}\]
subject to
\[\label{iccommhigjh}\tag{$IC_1$} -p b^2 - \left(1-p\right)\left(1 + b\right)^2 \geq -q b^2 - \left(1-q\right)\left(1+b\right)^2 - c\text{ ,}\]
and
\[\label{iccommlojw}\tag{$IC_0$} -q\left(1-b\right)^2 - \left(1-q\right)b^2 \geq -p\left(1-b\right)^2 - \left(1-p\right)b^2 - c\text{ .}\]

By our parametric assumptions Constraint \ref{iccommhigjh} does not bind, and so the optimum is given by a binding Constraint \ref{iccommlojw} and either \(q = 0\) or \(p = 1\), the latter if and only if \(\mu \geq 1/2\). Thus,

\begin{remark}\label{twoactremark}Unless \(\mu = 1/2\), the optimal signal is unique. It is given as follows:
    \begin{enumerate}[label={(\roman*)},noitemsep,topsep=0pt]
        \item If \(\mu > 1/2\), \(p^* = 1\) and \(q^* = 1 - c/\left(2b - 1\right)\).
        \item If \(\mu < 1/2\), \(p^* = c/\left(2b-1\right)\) and \(q^* = 0\).
        \item If \(\mu = 1/2\), \(p^*\) ranges from \(c/\left(2b-1\right)\) to \(1\) and \(q^* = p^* - c/\left(2b-1\right)\).
    \end{enumerate}
\end{remark}
%Unsurprisingly, as the sender's bias (\(b\)) increases or as his lying penalty (\(c\)) decreases, the receiver obtains less information in a Blackwell (\cite{blackwell}) sense. When there is zero cost to lying (\(c = 0\)), the receiver can elicit no information.

Now suppose $A = \left\{0, \hat{a}, 1\right\}$, where \(\hat{a} \in \left(2b-1,1\right)\). In the receiver's commitment problem, she commits to four conditional distributions over actions $p \coloneqq \mathbb{P}\left(\left.1\right|1\right)$, $\hat{p} \coloneqq \mathbb{P}\left(\left.\hat{a}\right|1\right)$, $q \coloneqq \mathbb{P}\left(\left.1\right|0\right)$ and $\hat{q} \coloneqq \mathbb{P}\left(\left.\hat{a}\right|0\right)$. Leaving the details to Appendix \ref{3actionsderivation}, we have:
\begin{remark}\label{3actions}Except for a knife-edge case, the commitment solution is unique. If the prior is sufficiently high, the commitment solution is \(q^{*} = \hat{p}^{*} = 0\), \(p^{*} = 1\), and \(\hat{q}^{*} = \frac{2b-c-1}{\hat{a}\left(2b- \hat{a}\right)}\). Otherwise, the commitment solution is \(q^{*} = \hat{p}^{*} = \hat{q}^{*} = 0\), and \(p^{*} = \frac{c}{2b-1}\).
\end{remark}
Notably, the commitment solution does not satisfy the obedience constraints when the prior is sufficiently high: when the receiver is recommended \(\hat{a}\) by the mechanism, she knows the sender is type \(0\) and is unwilling to follow the recommendation. Furthermore, the explicit expression for a ``sufficiently high'' prior is 
\[\frac{\mu}{1-\mu} \geq \frac{\left(2b-1\right)\hat{a}}{2b-\hat{a}}\text{.}\]
The left-hand side of this is strictly increasing in \(b\), so we see that commonality of interest (a smaller \(b\)) need not make the commitment solution easier to obtain via opacity design. If the prior is not ``sufficiently high'' (the stated inequality does not hold), the commitment solution satisfies the obedience constraints, so the receiver can obtain the commitment solution with the weaker form of commitment corresponding to opacity.
%\(\hat{q} = \left(2 b - 1 - c\right)/\left(2 b \hat{a} - \hat{a}^2\right)\)

This exercise illustrates that the finding of Proposition \ref{bigiftrue2} need not persist if the receiver has three or more actions.\footnote{The setting inhabited by \cite{ball} is another in which commitment is better for the receiver than opacity.} What is it that makes the two-action case special? The crucial aspect of the receiver's binary decision is that, following a recommendation by the signal, there is \textit{only one way in which she can be disobedient}. Consequently, disobeying a recommendation would yield her the same payoff from an \textit{ex ante} perspective as if there were no information transmitted whatsoever, since she would be taking just one action. However, she can always trivially obtain that payoff by choosing a completely uninformative signal. Thus, disobedience (of the signal corresponding to the commitment solution) must not be profitable.

In the commitment solution, when \(\mu\) is high, it is important for the receiver to take the correct action when the state is \(1\), as it occurs so frequently. In order to maintain incentive compatibility for the sender in state \(0\), this requires taking a non-zero action with positive probability. Moreover, it is always best (with commitment) to take action \(\hat{a}\). This cannot be done in the opacity design problem; however, as it exposes the low state.% \(T\), the maximal number of actions that guarantee opacity \(=\) commitment (Corollary \ref{maximalcorr}), is \(2\).
%When \(\mu\) is low, it is less important to take the correct action when the state is \(1\), and the receiver never takes action \(\hat{a}\). 

%Furthermore, note that in the commitment solution, outside of the parameter region in which the receiver randomizes, only two actions are chosen. This suggests a potentially interesting avenue for future research. \cite{szalay} illustrates how a principal can restrict her action set to only extreme actions in order to incentivize an agent to acquire information. Here, we discover another benefit to restricting actions: the receiver can attain her commitment solution by judiciously restricting her set of actions\footnote{There is historical evidence of such behavior: refer, e.g., to Odysseus lashing himself to the mast in order to hear the sirens.} before committing to a signal, provided she and the sender are not too dissimilar.

\subsection{A Continuum of Actions}\label{mtbgame}
Now let $A = \left[0,1\right]$.

\subsubsection{Commitment}
Suppose first the receiver can commit to a distribution over actions as a function of the sender's message. Formally, the receiver commits to two conditional distributions over actions $P\left(a\right) \coloneqq \mathbb{P}\left(A \leq \left.a\right|1\right)$ and $Q\left(a\right) \coloneqq \mathbb{P}\left(A \leq \left.a\right|0\right)$. Appealing to the revelation principle, the receiver solves 
\[\max_{P, Q}\left\{-\mu \int_{0}^{1}\left(a-1\right)^2dP\left(a\right) - \left(1-\mu\right) \int_{0}^{1}a^2dQ\left(a\right)\right\}\text{ ,}\]
subject to
\[\label{iccommhigh}\tag{$IC_1$} -\int_{0}^{1}\left(a-1-b\right)^2dP\left(a\right) \geq -\int_{0}^{1}\left(a-1-b\right)^2dQ\left(a\right) - c\text{ ,}\]
and
\[\label{iccommlow}\tag{$IC_0$} -\int_{0}^{1}\left(a-b\right)^2dQ\left(a\right) \geq -\int_{0}^{1}\left(a-b\right)^2dP\left(a\right) - c\text{ .}\]
Absent the IC conditions, the receiver would choose a  degenerate $P$ and $Q$ on $1$ and $0$, respectively. However, this is not incentive compatible: the sender prefers to lie in state $0$. Despite this, this problem is not too difficult. We introduce the following function of the bias $b$:
\[\gamma\left(b\right) \coloneqq 2b-1-b^{2}\mu\left(2-\mu\right) \text{ ,}\] and state the following proposition, leaving its derivation to the appendix.
\begin{proposition}\label{mtscomm}
If the prior is sufficiently high, $c \geq \gamma\left(b\right)$, in the optimal commitment solution, both $P$ and $Q$ are degenerate distributions. Otherwise, $Q$ is degenerate but $P$ is binary, with support on $\left\{0,1\right\}$.
\end{proposition}
This proposition is similar in spirit to Remark \ref{3actions}. There is a tension the receiver must resolve in choosing $P$ and \(Q\). When \(\mu\) is high, it is relatively more damaging to randomize (which hurts the receiver due to the strict concavity of her objective) in state \(1\). Accordingly, the receiver prefers a degenerate distribution, even though it makes deviating attractive for the sender in state $0$.

In this game, the receiver does strictly better by having a strong form of commitment to actions instead of only information design:
\begin{corollary}\label{notcomm}
In this game opacity $\neq$ commitment.
\end{corollary}
\subsubsection{Opacity Design}\iffalse
When there are more than two actions, opacity does not equal commitment in general (and does not in this section's problem under study). When it does not hold, how are we to solve the opacity design problem? Any strategy vector $\sigma$ by the sender and signal $\pi$ yields a distribution over posteriors $F \in \Delta\Delta\left(\Theta\right)$ that must be Bayes-plausible. That is, denoting a posterior by $x = x\left(\theta\right) \in \Delta\left(\Theta\right)$, $\mathbb{E}\left[x\right] = \mu$. Given this, we can directly wrap in the obedience constraints for the receiver and write the receiver's reduced form payoff as a function of her posterior $x$: 
\[v_{R}\left(x\right) \coloneqq \max_{a \in A}\mathbb{E}_{x}\left[u_{R}\left(a,\theta\right)\right] \text{ .}\]
We can also write the sender's reduced form payoff as a function of the receiver's posterior $x$ and his message $m$ as
\[v_{S}\left(x,m\right) \coloneqq u_{S}\left(a,\theta,m\right)\]

$F$ also determines the conditional distributions over posteriors for each on-path message $m$\fi

Because of the quadratic loss specification, the obedience constraint can be folded into the optimization problem in a concise way. In particular, given any posterior belief $x \coloneqq \mathbb{P}\left(1\right)$, the receiver's optimal action is merely $x$, yielding a payoff (as a function of the belief) $-x\left(1-x\right)$. It is again without loss of generality to restrict the sender to truthful ``reports'' of the state, and the receiver's choice variable is any distribution $F$ in the set of Bayes-plausible distributions $\mathcal{F}\left(\mu\right)$. Naturally, the sender's reports must be incentive compatible, so the receiver solves
\[\max_{F \in \mathcal{F}\left(\mu\right)}\left\{-\int_{0}^{1}x\left(1-x\right)dF\left(x\right)\right\} \text{ ,}\]
subject to 
\[\label{icidhigh}\tag{$IC_1$}- \int_{0}^{1}\left(x-b-1\right)^2\frac{x}{\mu}dF\left(x\right)dx \geq - \int_{0}^{1}\left(x-b-1\right)^2\frac{1-x}{1-\mu}dF\left(x\right)dx - c\text{ ,}\]
and
\[\label{icidlow}\tag{$IC_0$}- \int_{0}^{1}\left(x-b\right)^2\frac{1-x}{1-\mu}dF\left(x\right)dx \geq - \int_{0}^{1}\left(x-b\right)^2\frac{x}{\mu}dF\left(x\right)dx - c\text{ .}\]
It is clear that the first constraint (as in the commitment problem) must be slack. Moreover, following \cite{dov} we can write the following Lagrangian corresponding to the receiver's \textit{minimization problem}:
\[\mathcal{L}\left(x\right) = x\left(1-x\right) + \lambda \left(-c + \frac{\left(\mu-x\right)\left(x-b\right)^2}{\left(1-\mu\right)\mu}\right) \text{ .}\]
Leaving the details (and a figure) to the appendix, we unearth the following proposition:
\begin{proposition}\label{infomts}
There is a uniquely optimal signal. It has two realizations and yields posteriors $\left\{\ubar{x},1\right\}$, where $\ubar{x} \in \left(0,\mu\right)$.
\end{proposition}
At the optimum, the sender in state $1$ is occasionally exposed, or else is pooled with the sender in state $0$ by the signal. The natural relationship between the conflict of interest and the informativeness of the signal established in Remark \ref{twoactremark} persists:
\begin{corollary}
As incentives become more aligned, or the cost of misrepresentation increases ($b$ decreases or $c$ increases, respectively), the optimal signal becomes more informative.
\end{corollary}
In this game, any lack of informativeness is due to the problem of providing incentives to the sender. As this concern lessens, the optimal signal adjusts to improve the receiver's decision-making.

A comparison between the opacity design solution in this game to the commitment solution is illuminating. In particular, observe that with opacity design the optimal signal is always stochastic: the sender’s message in state $1$ is occasionally mixed with the sender’s message in state $0$. In contrast, in the commitment solution, it is only when incentives are not aligned and the cost of misrepresentation is low that the receiver’s actions are random. With opacity design, the recommendations by the signal must be random, in order for the receiver to be willing to obey its directives. There is no such issue with commitment to actions, randomization is only needed as a direct incentive device that mechanically makes a deviation by the sender in state $0$ less enticing.

%Finally, this 

\section{The Rewards to \textit{ex ante} Learning}\label{convexity}

We established above that in two-action games opacity design is as good as commitment. Its usefulness is not limited merely to the game itself. As we discover in this section, it also guarantees that the value of \textit{ex ante} information is always (weakly) positive.

Let us revisit the ``Beer-Quiche'' example. Suppose that \textit{ex ante}, prior to the signaling game, the receiver has a chance to acquire public information. She may for instance scrutinize the sender's clothing, or the company that he keeps, both of which are correlated with the state. Must such \textit{ex ante} information acquisition benefit the receiver?

We focus on two cases: that in which the receiver may design the opacity in the ensuing game, and that in which she may not. Recall that the opacity design payoff is $V = V(\mu)$. We denote the receiver's payoff without opacity design as $V^{T} = V^{T}(\mu)$.\footnote{Formally, the \textbf{Payoff Without Opacity Design} is the receiver's maximal payoff in the receiver-optimal PBE of the signaling game without opacity design when the prior is $\mu$.} We wish to discover whether these functions are convex. Indeed, the convexity of these functions is equivalent to the value of information being (weakly) positive--any secant line lies above the function so any splitting of a belief must yield an (expected) payoff that is weakly higher than the payoff at the belief.

In the example, it is straightforward to verify that $V$ and $V^{T}$ are
\[V(\mu) = \begin{cases}
1-\frac{\mu}{2}, & \quad \mu \leq \frac{1}{2}\\
\frac{1+\mu}{2}, & \quad \mu \geq \frac{1}{2}
\end{cases}, \qquad \text{and} \qquad V^{T}(\mu) = \begin{cases}
1-\mu, & \quad \mu \leq \frac{1}{2}\\
\mu, & \quad \mu \geq \frac{1}{2}
\end{cases} \text{ ,}\]
which are convex. Hence, regardless of whether the receiver can design the opacity, \textit{ex ante} learning is always (at least weakly) beneficial. As we will see shortly, in two-action games \textit{ex ante} information acquisition is always good, provided the receiver can design the opacity in the ensuing game. If she cannot, then information up front may actually hurt the receiver. That is, there is information that she would refuse, even if it were free.

Formally, we model \textit{ex ante} information acquisition or fact-finding as follows. Fix a signaling game, and suppose that prior to participating in the game, the receiver may acquire public information. Initially, the receiver and the sender share some prior, $\mu_{0}$, and there is some finite (or at least compact) set of signal realizations $Y$ and a signal, $\zeta\colon \Theta \to \Delta(Y)$, whose realization is public. Call $\zeta$ the \hypertarget{ie}{\textbf{Initial Experiment}}. This begets a distribution over posteriors, where the posterior following signal realization $y$ is $\mu_{y}$. The sender and receiver then take part in the signaling game, where the common prior for the game is now $\mu_{y}$. 

Obviously, regardless of the game, there are some experiments that benefit the receiver (or at least do not hurt her): for instance a fully informative initial experiment always benefits the receiver; but surprisingly, there are also experiments that hurt the receiver.

\begin{example} There are four states, $\Theta = \left\{\theta_{1}, \theta_{2}, \theta_{3}, \theta_{4}\right\}$, and a belief is a quadruple $(\mu_{1}, \mu_{2}, \mu_{3},\mu_{4})$, where $\mu_{i} \coloneqq \Pr\left(\Theta=\theta_{i}\right)$ for all $i = 1, 2, 3, 4$ and $\mu_{1} + \mu_{2} + \mu_{3} + \mu_{4} = 1$.

The belief can be fully described with just three variables; hence, depicting the receiver's payoff as a function of the belief requires four dimensions. The current (two-dimensional) medium of this paper renders this impossible, so instead we restrict attention to a family of experiments that involve learning on just one dimension. That is, we fix $\mu_{1} = 1/3$ and $\mu_{3} = 1/8$, and consider only the receiver's payoff as a function of her (prior) belief about states $\theta_{2}$ and $\theta_{4}$. Learning is on just one dimension, and so (abusing notation) we rewrite the receiver's belief $\mu_{2}$ as $\mu$ and $\mu_{4}$ as $13/24 - \mu$, where $\mu \in [0,13/24]$.

In states $\theta_{1}$ and $\theta_{2}$, action $a_{2}$ is the correct action for the receiver; and in states $\theta_{3}$ and $\theta_{4}$, action $a_{1}$ is correct:

\begin{center}
\begin{tabular}{ |c|c|c|c|c| } 
\hline
Action & $\theta_{1}$ & $\theta_{2}$ & $\theta_{3}$ & $\theta_{4}$ \\ 
$a_{1}$ & $0$ & $0$ & $1$ & $2$ \\ 
$a_{2}$  & $1$ & $1$ & $0$ & $0$ \\ 
\hline
\end{tabular}
\end{center}

\medskip

Likewise, the sender's state-dependent payoffs from message, action pairs are

\medskip

\begin{center}
\begin{tabular}{ |c|c|c|c|c| } 
\hline
state & $\theta_{1}$ & $\theta_{2}$ & $\theta_{3}$ & $\theta_{4}$ \\
\hline
message & $m_{1}$ \quad $m_{2}$ \quad $m_{3}$ & $m_{1}$ \quad $m_{2}$ \quad $m_{3}$ & $m_{1}$ \quad $m_{2}$ \quad $m_{3}$ & $m_{1}$ \quad $m_{2}$ \quad $m_{3}$\\
\hline
$a_{1}$ & $1$ \quad $0$ \quad $0$ & $0$ \quad $1$ \quad $0$ & $0$ \quad $1$ \quad $3$ & $-1$ \quad $2$ \quad $-2$\\
\hline
$a_{2}$ & $1$ \quad $0$ \quad $0$ & $0$ \quad $1$ \quad $0$ & $2$ \quad $4$ \quad $0$ & $5/4$ \quad $0$ \quad $-1$\\
\hline
\end{tabular}
\end{center}

\medskip

\begin{figure}
    \centering
    \includegraphics[scale=.4]{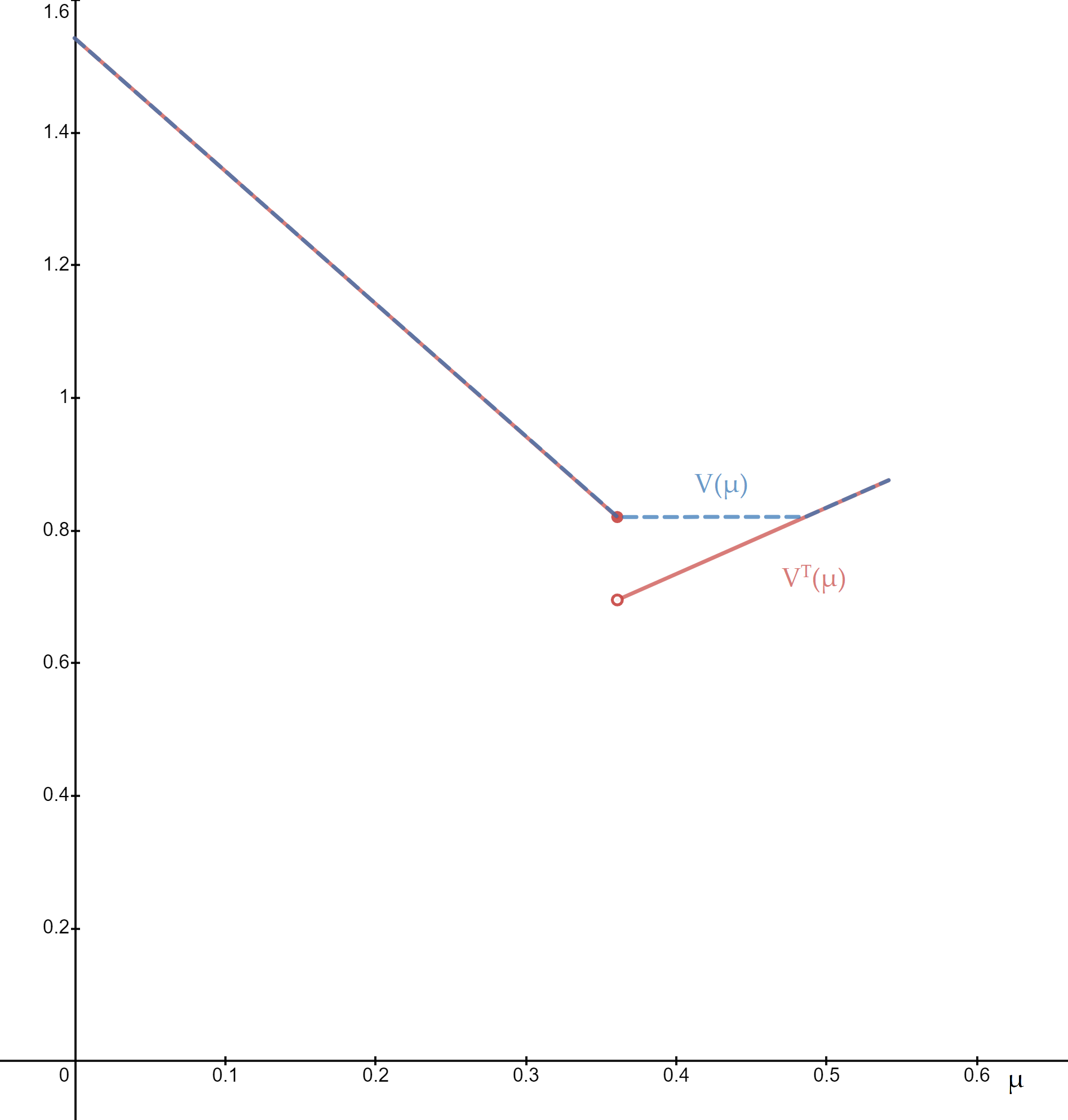}
    \caption{Payoffs With and Without Opacity Design}
    \label{4dconvex}
\end{figure}

In Figure \ref{4dconvex} we depict the opacity design payoff ($V$) and the payoff without opacity design ($V^{T}$). Explicitly, those functions are
\[V(\mu) = \begin{cases}
\frac{37}{24}- 2\mu, & \quad \mu \leq \frac{13}{36}\\
\frac{59}{72}, & \quad \frac{13}{36} \leq \mu \leq \frac{35}{72}\\
\frac{1}{3} + \mu, & \quad \frac{35}{72} \leq \mu \leq \frac{13}{24}
\end{cases}, \qquad \text{and} \qquad V^{T}(\mu) = \begin{cases}
\frac{37}{24}- 2\mu, & \quad \mu \leq \frac{13}{36}\\
\frac{1}{3} + \mu, & \quad \frac{13}{36} < \mu \leq \frac{13}{24}
\end{cases} \text{ .}\]

Without opacity design, \textit{ex ante} information acquisition can hurt the receiver by eliminating particularly lucrative equilibria. In this example, if $\mu \leq 13/36$, there is an equilibrium with partial separation, which provides some useful information to the receiver. On the other hand, should $\mu$ strictly exceed $13/36$, the only equilibria beget the pooling payoff--the receiver becomes too sure that the state is not $\theta_{3}$ or $\theta_{4}$. Thus, no useful information is transmitted, to the receiver's detriment. 

As Figure \ref{4dconvex} illustrates, the issue engendered by \textit{ex ante} learning--that the resulting belief may beget a strictly worse equilibrium in the signaling game--is ameliorated if the receiver can design the game's opacity. The optimal $\pi$ allows for a moderately informative equilibrium on the interval $[13/36,35/72]$, to the receiver's benefit.
\end{example}

This counterexample suffices to prove the following proposition:

\begin{proposition}\label{410}
In signaling games with four or more states, three or more messages, and two actions, the receiver's payoff without opacity design is not generally convex in the prior.
\end{proposition}

Although the example illustrates that learning may be injurious when the receiver is unable to design the opacity in the game that follows, it also suggests that things may be different if the receiver can design the opacity in the ensuing game. In the example, opacity design renders the value of \textit{ex ante} information positive. As the next pair of results illustrate, this result is general.

With commitment power, information acquisition is always beneficial:

\begin{lemma}\label{12thm}
The commitment payoff is convex in the prior, $\mu$.
\end{lemma}

Recall the earlier result--that opacity equals commitment. Intuitively, the receiver's commitment problem is simply a decision problem, once the sender's best response (to the commitment) is folded into the objective. Thus, Proposition \ref{bigiftrue2} and Lemma \ref{12thm} combine to yield  

\begin{proposition}\label{convconv}
In two-action signaling games, the opacity design payoff is convex in the prior, $\mu$.
\end{proposition}

Not only is opacity design as good as commitment in binary-action games, it also guarantees that \textit{ex ante} learning can only benefit the receiver, thus revealing a second avenue through which opacity design is useful. Furthermore, just as Proposition \ref{bigiftrue2} can be extended to allow for more than one sender, so can Proposition \ref{convconv}. \textit{Viz.}, in binary-action multi-sender signaling games, the value of (public) information is always positive for the receiver, provided the receiver can commit to an information structure. Here is an easy argument: at initial belief $\mu_0$, let $\pi^{*}\colon M \to \Delta\left\{a_1,a_2\right\}$ be an optimal commitment strategy for the receiver and let $V^{*}$ be the receiver's payoff from such a strategy. Following any realization of the initial experiment it is obvious that receiver can choose the same $\pi^{*}$ as it did at the initial belief. Thus, the receiver can do no worse than $V^{*}$ in expectation (with respect to $\zeta$), hence we have convexity. That opacity equals commitment completes the reasoning.

Even without opacity design, in two-action cheap-talk games \textit{ex ante} information cannot hurt the receiver:
\begin{proposition}\label{main1}
In two-action cheap-talk games, the value of information is always positive for the receiver
\end{proposition}

This proposition is closely related to Proposition \ref{bigiftrue2}. It is easy to see that at any prior, there must be a receiver-optimal equilibrium in which at most two messages are sent--akin to the action recommendations of an optimal signal. Then, just as the opacity equals commitment result follows from the fact that disobedience cannot be interim optimal--since the receiver could always have just committed to take the same action in the first place--the value of \textit{ex ante} information must be optimal in binary-action cheap-talk games even without opacity design. At any new prior for the game, either the same strategy vector for the sender is part of the receiver optimal equilibrium--the receiver is still willing to follow the ``recommendations''--or it isn't. But if it isn't, the receiver strictly prefers to choose the same action after each message instead of following the ``recommendations.'' An equilibrium that yields this payoff always exists--for instance, the babbling equilibrium--and so the value of initial learning must be positive.

Beyond cheap talk, it is easy to identify other (strong) sufficient conditions that guarantee a positive value of information in signaling games without opacity design. Indeed, any conditions that engender a fully separating equilibrium ensure a positive value of information, since in such equilibria the receiver's posterior beliefs do not depend on her prior. In the online appendix, we specify further nontrivial sufficient conditions.

%This result relies heavily on the number of (on-path) messages to which we may limit the sender in a receiver-optimal equilibrium. When there are just two states, the belief space is one-dimensional, which constrains the effect of information and ensures that there is a receiver-optimal equilibrium in which two messages (or fewer) are used. With two actions, the costless nature of cheap talk messages also ensures that there is an optimal equilibrium that requires at most two messages. This observation is then be used in conjunction with either the binary state or binary action nature of the game to establish that the receiver's payoff 

%the equilibrium play following each realization  must yield the receiver a payoff at least as high as equilibrium play as if the equilibrium play engendered a distribution over posteriors that was more informative 

\subsection{Related Analyses of the Value of Information}

The closest work to this portion of the paper is \cite{chen}, who shows that in a two state quadratic cheap talk setting, the value of information may be negative for the receiver.\footnote{Note that her result does not contradict Proposition \ref{main1} below--in her setting the sender, himself, is only partially informed about the state and thus learns from the public signal as well.} Also similar is \cite{mb}, who shows that the value of \textit{private} information for the receiver in the uniform-quadratic setting of \cite{cs} may be negative. \cite{wekslerzik} introduce testing into a signaling environment, wherein a receiver chooses a test (whose realization is private) which the sender observes before sending a costly a signal. Notably, the receiver may prefer less informative tests. The rationale is similar to the results on the potential negative value of information of this paper--the additional information provided by the test may lead to a worse equilibrium. In contrast, \cite{lichtig2020info} show that increased information for the \textit{sender} benefits the receiver in a general class of voluntary disclosure games.

There are also a number of other papers, including \cite{kamien,goos,bass,teacher,pes,meyer}; and \cite{ui}, that look at the value of information in games. However, none of these explore the value of information in the same sense as this paper. Here, we are interested in something much closer to a decision problem for the receiver, with the \textit{caveat} that the information generation process is endogenous and generated by equilibrium play of the sender. Outside of that there are no strategic concerns, and the sender is perfectly informed, so there is no learning on his part. Consequently, this work's (and Chen's) analysis is more similar in spirit to e.g. \cite{blackwell, blackwell2}; and \cite{ram}, who explore the value of information in decision problems.

\section{Discussion}\label{conc}

Opacity design, in two-action signaling games, enables the receiver to achieve the same payoff as in the case wherein she had the power of committing \textit{ex ante} to actions conditioned on the signal choice of the sender. This is both unexpected, that for a broad class of games information design is as powerful as commitment; and useful, since the receiver's commitment problem is merely a linear program. 

Also surprising is that even in games with just two actions, \textit{ex ante} learning may hurt the receiver, provided she cannot design the opacity in the game. The resulting drop in (useful) information provided in equilibrium at the new priors may be so great so as to overpower the initial gain in information. On the other hand, if she may design the opacity, the value of information is always positive--the receiver always benefits from information, no matter the form it takes.

One final comment: there is recent experimental evidence that senders respond to different levels of opacity, which advocates for the usefulness of this paper's results. Specifically, \cite{blume2} find that senders do respond as predicted to the introduction of noise in a communication setting. Relatedly, the randomized response technique of \cite{warner}, in which a survey respondent's answers are garbled (ostensibly to preserve anonymity), is in similar spirit to the information design in signaling games proposed by this work. The success of that technique--refer, e.g., to \cite{blume1}, who show experimentally that randomized response can induce senders to be significantly
more truthful--provides suggestive evidence that the information design we studied may be useful in practice. We leave this for future work.

%Thus, Proposition \ref{bigiftrue2} is not merely interesting but perhaps also practical, since it makes the design of optimal information structures so easy.

\appendix

\section{Omitted Proofs}

\subsection{Remark \ref{revp} Proof}
\begin{proof}
Consider any equilibrium triple $(\pi, \sigma, \rho)$. Now introduce for each signal realization $x_e$ and action $a_{l}$ in the support of $\rho(\cdot|x_{e})$ two new mappings:
\begin{enumerate}[noitemsep,topsep=0pt]
    \item $\hat{\pi}\colon M \to \Delta\left(\hat{A}\right)$, where $\hat{\pi}(a_{l}^{e}|m_{j}) \coloneqq \pi(x_{e}|m_{j})\rho(a_{l}|x_{e})$; and
    \item $\hat{\rho}$, where $\hat{\rho}(a_{l}|a_{l}^{e}) = 1$ and $\hat{\rho}(a_{l}|a_{l'}^{e}) = 0$ for all $l' \neq l$.
\end{enumerate}
That is, $a^e_l$ is the instruction to play $a_l$ that induces the same belief as $x_e$. Clearly set $\hat{A}$ may be larger than $A$--it may have multiple ``duplicate'' recommendations.

By construction, for each $a_{l}^{e}$, action $a_{l}$ is a best response. Moreover, both the obedience and IC constraints are satisfied, and the expected payoff for the receiver is the same. Finally, introduce garbling $g\colon \hat{A} \to \Delta \left(A\right)$, where $g(a_{l}|a_{l'}^{e}) = 1$ for $l = l'$ and $g(a_{l}|a_{l'}^{e}) = 0$ for $l \neq l'$. 

Define $\pi' \coloneqq g \circ \hat{\pi}$. It is easy to see that the IC constraints remain satisfied (since each message will lead to the same distribution of actions chosen by the receiver). Moreover, the obedience constraints must be satisfied as well since $\pi'$ is less (Blackwell) informative than $\hat{\pi}$. Hence $(\pi', \sigma', \rho')$ is also an equilibrium, where $\rho'(a_{l}|a_{l'}) = 1$ for $l = l'$, $\rho'(a_{l}|a_{l'}) = 0$ for $l \neq l'$, and $\sigma' = \sigma$; and the receiver's expected payoff remains the same.
\end{proof}

\iffalse
\subsection{Corollary \ref{maximalcorr} Proof}
\begin{proof}
    If \(\left|A\right| = 1\), the result is trivially true. Proposition \ref{bigiftrue2} states that if \(\left|A\right| = 2\), opacity \(=\) commitment. Thus, if \(\left|A\right| \leq 2\) opacity \(=\) commitment, which implies the result.\end{proof}
    \fi

\subsection{Remark \ref{3actions} Proof}\label{3actionsderivation}
\begin{proof}
    The receiver solves
\[\max_{p, \hat{p}, q, \hat{q}}\left\{-\mu \left(\hat{p}\left(1-\hat{a}\right)^2 + 1 - p - \hat{p}\right) - \left(1-\mu\right)\left(q + \hat{q}\hat{a}^2\right)\right\}\text{ ,}\]
subject to
\[\label{iccommhigjjh}\tag{$IC_1$} -p b^2 - \hat{p}\left(\hat{a} - 1 - b\right)^2 - \left(1-p-\hat{p}\right)\left(1 + b\right)^2 \geq -q b^2 - \hat{q}\left(\hat{a} - 1 - b\right)^2 - \left(1-q-\hat{q}\right)\left(1 + b\right)^22 - c\text{ ,}\]
and
\[\label{iccommlojjw}\tag{$IC_0$} -q\left(1-b\right)^2 - \hat{q}\left(\hat{a}-b\right)^2 - \left(1-q- \hat{q}\right)b^2 \geq -p\left(1-b\right)^2 - \hat{p}\left(\hat{a}-b\right)^2 - \left(1-p- \hat{p}\right)b^2 - c\text{ .}\]
It is straightforward to verify that \ref{iccommhigjjh} does not bind. On the other hand, \ref{iccommlojjw} must hold with equality at the optimum. Furthermore, we must have \(p + \hat{p} \leq 1\) (in addition to \(p, \hat{p}, q, \hat{q} \in \left[0,1\right]\) and \(q + \hat{q} \leq 1\)). Let us first guess that \(p + \hat{p} = 1\). We can substitute in for \(\hat{p}\) into \ref{iccommlojjw}, and rearrange to get \(p\) as a function of \(q\) and \(\hat{q}\):
\[p = \frac{\left(2b-1\right)q+2\hat{a}b\left(\hat{q}-1\right)-\hat{a}^2\left(\hat{q}-1\right)+c}{\left(\hat{a}-1\right)\left(-2b+\hat{a}+1\right)}\text{.}\]
As \(0 \leq p \leq 1\), this becomes
\[\frac{2bq-q+c-2b+1}{\hat{a}\left(\hat{a}-2b\right)}\le \hat{q} \leq \frac{2bq-q+c-2\hat{a}b+\hat{a}^{2}}{\hat{a}\left(\hat{a}-2b\right)}\text{.}\]
We also substitute for \(p\) and \(\hat{p}\) into the objective to obtain a two-variable linear program, which we solve for \(q\) and \(\hat{q}\). All in all, we obtain 
\(q = \hat{p} = 0\), \(p = 1\), and \[\hat{q} = \frac{2b-c-1}{\hat{a}\left(2b- \hat{a}\right)}\text{,}\]which produces a payoff for the receiver of 
\[\tag{\(A_1\)}\label{v1}-\frac{\left(1-\mu\right)\hat{a}\left(2b-1-c\right)}{2b-\hat{a}}\text{.}\]
It is also easy to check that at the optimum we cannot have \(q + \hat{q} = 1\). This leaves only--as there exists a solution to a linear program at a vertex of the constraint polytope--three of \(p, \hat{p}, q\), and \(\hat{q}\) equaling \(0\); and the fourth pinned down by those and the binding \ref{iccommlojjw} as a solution. Checking them one by one, we see that only \(q = \hat{q} = \hat{p} =0 \) and
\[p = \frac{c}{2b-1}\text{,}\]
is ever optimal. It produces a payoff of 
\[-\frac{\left(2b-1-c\right)\mu}{2b-1}\text{.}\] Comparing it to Expression \ref{v1} yields the first part of the result. Finally, if the solution is not unique, then there must be multiple optimal solutions at vertices of the constraint polytope. However, this is only true if the parameters satisfy
\[\frac{\mu}{1-\mu} = \frac{\left(2b-1\right)\hat{a}}{2b-\hat{a}}\text{,}\] which is a knife-edge case. \end{proof}

\subsection{Proposition \ref{mtscomm} Proof}\label{mtscommproof}
\begin{proof}
It is clear that Constraint \ref{iccommhigh} must be slack. Therefore, by the concavity of $-a^2$ and $-\left(a-b\right)^2$, it is optimal for $Q$ to be degenerate on some action $y \in \left[0,1\right]$. Now let us define $v$ as the expectation of $P$. Given this, the (necessarily) binding Constraint \ref{iccommlow} can be rewritten as
\[\int_{0}^{1}a^2dP\left(a\right) = 2 b v - 2 b y - c + y^2 \text{ .}\] 
Substituting this into the objective and reducing (and eliminating constants), we obtain the following maximization problem
\[\max_{y, v \ \in \left[0,1\right]}\left\{2 \mu v\left(1-b\right) + 2 b \mu y - y^2\right\} \text{ ,}\]% - \mu\left(1-c\right)
subject to 
\[\tag{$A2$}\label{chainchain} v^2 \leq 2 b v - 2 b y - c + y^2 \leq v \text{ ,}\]
where this chain of inequalities follows from the bounds on the possible variance of $P$. Ignoring this constraint, we obtain that $v = 1$ and $y = b \mu$. However, this solution always violates one of the inequalities except in the knife-edge case where $c = 2b-1-b^{2}\mu\left(2-\mu\right)$. The result follows immediately. \end{proof}
\subsection{Corollary \ref{notcomm} Proof}
\begin{proof}
If both $P$ and $Q$ are deterministic, then opacity $=$ commitment if and only if $v = y$. Moreover, we can substitute this into Inequality \ref{chainchain}, which yields $v^2 - c \geq v^2$, a contradiction. If $P$ is not deterministic, the result is obvious.
\end{proof}

\subsection{Proposition \ref{infomts} Proof}
\begin{figure}
    \centering
    \includegraphics[scale=.3]{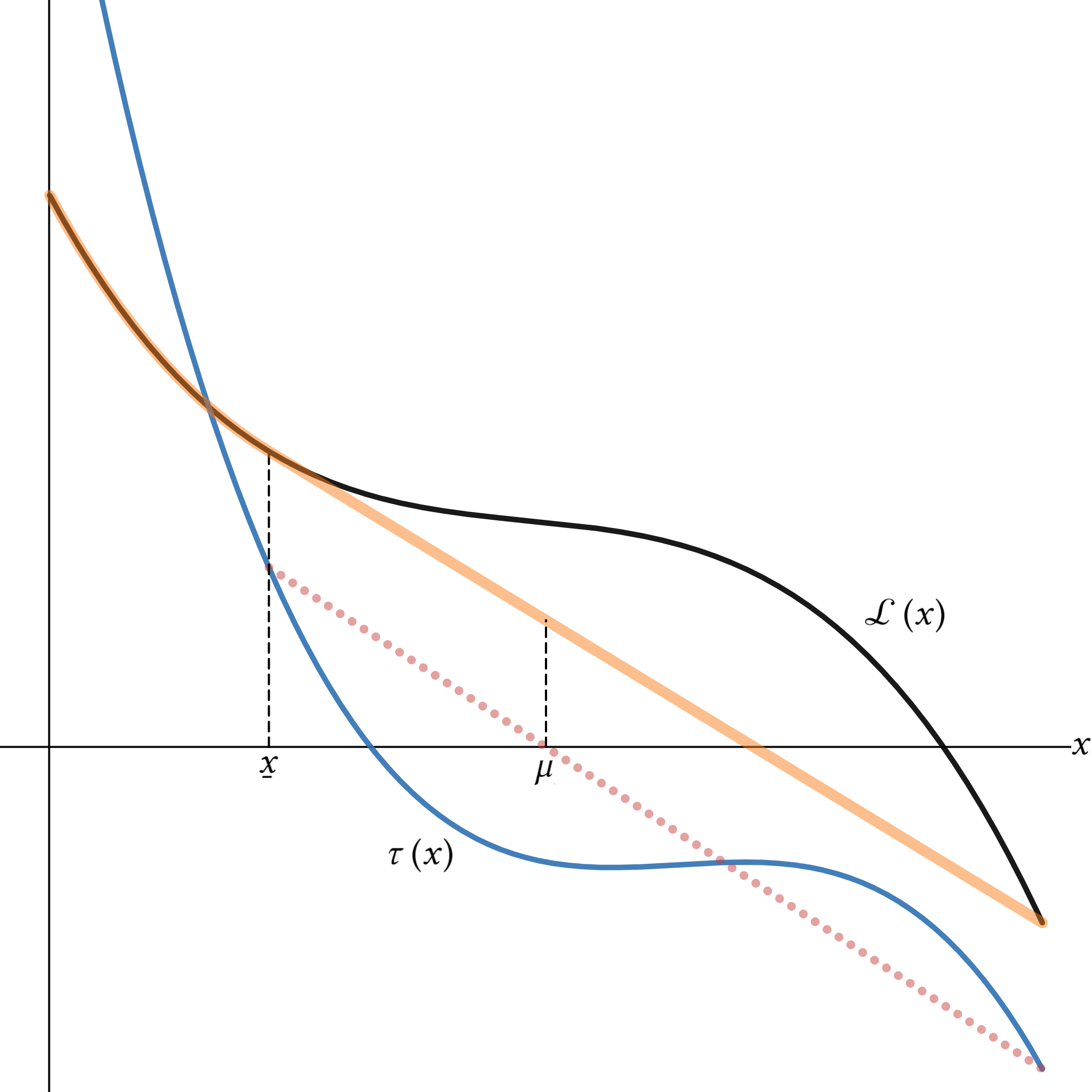}
    \caption{The Opacity Design Problem and its Solution}
    \label{lagrangian}
\end{figure}

\begin{proof}
A direct computation reveals that 
\[\mathcal{L}\left(x\right) = x\left(1-x\right) + \lambda \left(-c + \frac{\left(\mu-x\right)\left(x-b\right)^2}{\left(1-\mu\right)\mu}\right) \text{ ,}\]
is concave on $\left[0,1\right]$ if $\lambda$ is sufficiently low or else is convex then concave. Any line tangent to $\mathcal{L}\left(x\right)$ at a point $a \in \left[0,1\right]$ has equation
\[\begin{split}
    t\left(x\right) \coloneqq &\left[\lambda\left(\frac{2\left(\mu-a\right)\left(a-b\right)}{\left(1-\mu\right)\mu}-\frac{\left(a-b\right)^2}{\left(1-\mu\right)\mu}\right)-2a+1\right]x\\ &-\frac{2\lambda a^{3}-\left(\left(\mu+2b\right)\lambda+\left(\mu-1\right)\mu\right)a^{2}+\left(c\left(\mu-1\right)+b^{2}\right)\mu\lambda}{\left(\mu-1\right)\mu}
\end{split} \text{ .}\]
%Directly, 
%\[t\left(1\right) - \mathcal{L}\left(1\right) = \left(a-1\right)^2-\dfrac{{\lambda}\cdot\left(a-1\right)^2\left(2a-\mu-2b+1\right)}{\left(\mu-1\right)\mu} \text{ .}\]
The equation $t\left(1\right) - \mathcal{L}\left(1\right) = 0$ rearranges to
\[a = a\left(\lambda\right) = \frac{\left(\mu+2b-1\right)\lambda+\left(\mu-1\right)\mu}{2\lambda} \text{ .}\]
This function is strictly increasing in $\lambda$ and ranges from $-\infty$ to $\frac{\mu+2b-1}{2}$. Now, let us look for the line that intersects the two points $\left(\mu,0\right)$ and $\left(1, -c + \frac{\left(\mu-1\right)\left(1-b\right)^2}{\left(1-\mu\right)\mu}\right)$. It is \[-\frac{\left(c\mu+b^{2}-2b+1\right)\left(\mu-x\right)}{\left(\mu-1\right)\mu} \text{ ,}\]
and it intersects the curve \[\tau\left(x\right) \coloneqq -c + \frac{\left(\mu-x\right)\left(x-b\right)^2}{\left(1-\mu\right)\mu} \text{ ,}\] at the point 
\[\ubar{x} \coloneqq \frac{-\sqrt{4b^{2}+\left(-4\mu-4\right)b+\mu^{2}+\left(4c+2\right)\mu+1}+2b+\mu-1}{2} \text{ .}\]
$\ubar{x}$ is strictly decreasing in $c$ and equals $0$ when $c = 2b-1$. Accordingly $\ubar{x} > 0$. Moreover, it is clear that $\ubar{x} < \frac{\mu+2b-1}{2}$. Thus, by the intermediate value theorem, there exists a $\lambda > 0$ such that $a = \ubar{x}$. 

We have established the result: the support of the optimal distribution is $\left\{\ubar{x}, 1\right\}$, which is given by the convexification of the objective for the receiver. In this objective, the Lagrange multiplier, $\lambda$, is an endogenous object pinned down by the binding constraint for the sender in state $0$ (necessarily, $\mathbb{E}_{F}\left[\tau\left(x\right)\right] = 0$).\footnote{This is why we solved for the line that goes through the point $\left(\mu,0\right)$.} Figure \ref{lagrangian} illustrates the Lagrangian (solid black), its convexification (thick orange), the curve $\tau\left(x\right)$ (solid blue), and the splitting corresponding to the optimal $F$ (dotted red). \end{proof}

\subsection{Proposition \ref{410} Proof and Payoff Function Derivation}\label{410proof}
We derive the receiver's payoff as a function of the belief $\mu$ through a pair of claims. First,
\begin{claim}
Without opacity design, for any prior $\mu > 13/36$, there exists no equilibrium in which a message is played that induces a belief such that action $a_{1}$ is strictly optimal.
\end{claim}
\textit{Proof.} We can exhaustively proceed through each message:
\begin{enumerate}[noitemsep,topsep=0pt]
    \item Suppose $a_{1}$ is strictly optimal following $m_{1}$. Then, $\theta_{4}$ must have support of his mixed strategy on $m_{1}$. That gives him a payoff of $-1$, so he can deviate profitably to $m_{2}$. 
    \item Suppose $a_{1}$ is strictly optimal following $m_{2}$. Both $\theta_{3}$ and $\theta_{4}$ must have support of their mixed strategies on $m_{2}$.  Moreover, so much of their support must be on $m_{2}$ that the receiver must choose $a_{2}$ following $m_{1}$ (which will always be chosen by $\theta_{1}$). Hence, $\theta_{3}$ can deviate profitably to $m_{1}$.
    \item Suppose $a_{1}$ is strictly optimal following $m_{3}$. Consequently, $\theta_{3}$ must have some support of his mixed strategy on $m_{3}$ (since $\theta_{4}$ will never choose $m_{3}$). Moreover, $\theta_{3}$ cannot be choosing a pure strategy since otherwise he would have a profitable deviation to $m_{2}$. Hence, $\theta_{3}$ must be mixing over $m_{3}$ and $m_{2}$ (since $3$ is strictly larger than either of $\theta_{3}$'s payoffs for $m_{1}$). The receiver must also mix following $m_{2}$, so as to leave $\theta_{3}$ willing to mix. In particular, the receiver must choose $a_{2}$ with probability $2/3$ following $m_{2}$ and $a_{1}$ with probability $1/3$. But note that $\theta_{4}$ must also have support on $m_{2}$, which message would thus yield it a payoff of $2/3$, which is less than $5/4$, the payoff he would get from deviating profitably to $m_{1}$. \hfill $\blacksquare$
\end{enumerate}
As a result $V^{T} = \frac{1}{3} + \mu$ for all $\mu > 13/36$. Second,
\begin{claim}
For any $\mu \leq 13/36$, the payoff without opacity design is $37/24 - 2\mu$.
\end{claim}
\begin{proof}
First, an equilibrium that begets such a payoff exists. In state $\theta_{1}$ the sender chooses $m_{1}$, in states $\theta_{2}$ and $\theta_{4}$ he chooses $m_{2}$, and in state $\theta_{3}$ he chooses $m_{3}$. Upon observing $m_{1}$, the receiver chooses $a_{2}$, and upon observing either $m_{2}$ or $m_{3}$ the receiver chooses $a_{1}$.

It is immediately evident that neither $\theta_{1}$ nor $\theta_{2}$ have profitable deviations, since they are choosing strictly dominant strategies. On path, $\theta_{4}$ obtains $2$, whereas his payoff from deviating is less than $2$. Finally, $\theta_{3}$ obtains $3$ following $m_{3}$ and less than $3$ following any other message. The receiver's equilibrium payoff is
\[\frac{1}{3} + \frac{1}{8} + 2\left(\frac{13}{24}-\mu\right) = \frac{37}{24} - 2\mu \text{ .}\]
This payoff is the opacity design payoff and so must be maximal.
\end{proof}
By Proposition \ref{bigiftrue2}, to obtain $V$ we need only solve the commitment problem. Thus, its derivation reduces to a simple (though tedious) linear program, which is omitted.

\subsection{Lemma \ref{12thm} Proof}\label{12thmproof}
\begin{proof}
Recall that from Lemma \ref{lemlem} it is without loss of generality in the receiver's commitment problem to restrict the sender to pure strategies. Hence, the receiver solves 
\[\max_{\pi, \mathbf{s}}\left\{\sum_{i=1}^{n}\mu(\theta_{i})\sum_{l=1}^{k}\pi\left(a_{l}|m_{i}\right)u_{R}\left(m_{i},a_{l},\theta_{i}\right)\right\} \text{ ,}\]
such that
\[\sum_{l=1}^{k}\pi\left(a_{l}|m_{i}\right)u_{S}\left(m_{i},a_{l},\theta_{i}\right) \geq \sum_{l=1}^{k}\pi\left(a_{l}|m_{i}'\right)u_{S}\left(m_{i}',a_{l},\theta_{i}\right) \text{ ,}\]
for all $\theta_{i}$, $m_{i}'$, where $\mathbf{s}$ is a vector of pure strategies chosen by the sender and message $m_{i}$ is the message chosen in state $\theta_{i}$ (of course it is possible that $m_{i} = m_{k}$ for $k \neq i$ if the sender chooses the same message in states $\theta_{i}$ and $\theta_{k}$). For a fixed vector of pure strategies this is a linear program.

Naturally, the payoffs of the game may be such that for some vector of pure strategies, there exists no signal that is incentive compatible. Define set $F$ as the set of all pairs of signals and strategy vectors that are incentive compatible, with element $f \coloneqq (\pi, s)$. For any $f \in F$, we have
\[V_{f}(\mu) = \sum_{i=1}^{n}\mu(\theta_{i})\sum_{l=1}^{k}\pi\left(a_{l}|m_{i}\right)u_{R}\left(m_{i},a_{l},\theta_{i}\right) \text{ ,}\]
which is linear, and hence convex, in $\mu$.

Furthermore, by definition, $V(\mu) = \max_{f}V_{f}(\mu)$. Since $V_{f}$ is convex, $\epi V_f$ is also convex. Then, $\epi V = \epi \max_{f}V_{f} = \cap_{f \in F}\epi V_{f}$. Since the intersection of convex sets is also convex, $\epi V$ is also convex, hence $V$ is convex.
\end{proof}

\subsection{Proposition \ref{main1} Proof}\label{409proof}
\begin{proof}
First, we observe that it is WLOG to impose that at most two messages are sent in the receiver-optimal equilibrium. Take any receiver-optimal equilibrium in which more than two messages are sent and both actions are taken with probability $1$ following at least one message. Let messages $m_1$ and $m_2$ be two such on-path messages, each of which is followed by a different action ($a_1$ and $a_2$, respectively). Suppose only those two messages are used by the sender ($m_1$ if the sender is in a state where previously he had sent a message that was followed by $a_1$ and $m_2$ otherwise), followed by the two actions, respectively. This leaves the sender's incentives and (useful) information content of the messages unchanged. Consequently, this is also an equilibrium and yields the same payoffs to both players. 

Given this, the initial experiment either results in a posterior at which the receiver is willing to take action $a_1$ after $m_1$ and $a_2$ after $m_2$, or not. In the former case, the result is immediate: the incentives for the sender are unchanged so it is as if (from the receiver's the perspective) the information was exogenous. In the latter case, the proof is not much more difficult. Suppose at the new belief the receiver is unwilling to take action $a_1$ following $m_1$. This means that the receiver prefers to always take $a_2$ (after each message) than the status quo. However, there is always a babbling equilibrium in cheap-talk games, so at this new belief, there is an equilibrium in which the receiver takes a single action, which must yield the receiver a weakly higher payoff than under the prior. \end{proof}

\bibliography{sample.bib}

\end{document}